\documentclass[onecolumn,preprintnumbers,amsmath,amssymb,superscriptaddress]{revtex4}

\usepackage{graphicx}
\usepackage{dcolumn}

\newcommand{\be}{\begin{equation}}
\newcommand{\ee}{\end{equation}}
\newcommand{\bea}{\begin{eqnarray}}
\newcommand{\eea}{\end{eqnarray}}
\newcommand{\bm}[1]{\mathbf{#1}}

\newcommand{\ra}{\rangle}
\newcommand{\me}[3]{\langle #1 | #2 | #3 \rangle}

\newcommand{\lp}{\left(}
\newcommand{\rp}{\right)}

\def \cG{{\cal G}}

\def \cH{{\cal H}}

\newcommand{\ty}[1]{\mbox{\tiny #1}}

\begin{document}

\title{Moir\' e bands in twisted double-layer graphene}

\author{R. Bistritzer and A.H. MacDonald}
\affiliation{Department of Physics, The University of Texas at Austin, Austin Texas 78712 USA}


\maketitle

{\bf
A moir\' e pattern is formed when two copies of a periodic pattern are overlaid with a relative twist.
We address the electronic structure of a twisted two-layer graphene system,
showing that in its continuum Dirac model the moir\' e pattern periodicity leads to moir\' e
Bloch bands.  The two layers become more strongly coupled and the Dirac velocity crosses
zero several times as the twist angle is reduced.
For a discrete set of magic angles the velocity vanishes, the lowest moir\' e band flattens, and
the Dirac-point density-of-states and the counterflow conductivity are strongly enhanced.}

Low-energy electronic properties of few layer graphene (FLG) systems are known\cite{BernalHexagonal1,Ando,CastroNeto,BernalHexagonal2,BernalHexagonal3,BernalHexagonal4,FLGOptical,FLGthermal} to be
strongly dependent on stacking arrangement.  In bulk graphite $0^{\circ}$
and $60^{\circ}$ relative orientations of the individual layer honeycomb lattices yield rhombohedral and Bernal crystals,
but other twist angles also appear in many samples\cite{rotatedGraphite}.
Small twist angles are particularly abundant in epitaxial graphene layers grown on SiC\cite{rotatedEpitaxial}, but
exfoliated bilayers can also appear with a twist, and arbitrary alignments between adjacent
layers can be obtained by folding a single graphene layer\cite{foldedBilayer,Biro}.

Recent advances in FLG preparation methods have attracted theoretical attention\cite{Santos,Shallcross,ShallcrossLong,Mele,FLGus,localization}
to the intriguing electronic properties of systems with arbitrary twist angles, usually focusing
on the two-layer case.  The problem is mathematically interesting because a bilayer forms a two-dimensional crystal only at a discrete set of commensurate rotation angles;
for generic twist angles Bloch's theorem does not apply microscopically and
direct electronic structure calculations are not feasible.  For twist angles larger than a few degrees the two layers are
electronically isolated to a remarkable degree, except at a small set of
angles which yield low-order commensurate structures\cite{ShallcrossLong,FLGus}.
As the twist angles become smaller, interlayer coupling strengthens, and the quasiparticle velocity at the Dirac point begins to decrease

Here we focus on the strongly coupled small twist angle regime. We construct a
low energy continuum model Hamiltonian that consists of three terms:
two single layer Dirac-Hamiltonian terms that account for the isolated graphene sheets,
and a tunneling term that describes
hopping between layers.
The Dirac Hamiltonian\cite{GrapheneReview} for a layer rotated by an angle $\theta$ with respect to a fixed coordinate system is
\be
h_{\bm{k}}\lp \theta \rp = - v k
\left(
  \begin{array}{cc}
    0 & e^{i(\theta_{\bm{k}}-\theta)} \\
    e^{-i(\theta_{\bm{k}}-\theta)}    & 0 \\
  \end{array}
\right),     \nonumber
\ee
where $v$ is the Dirac velocity, $k$ is momentum magnitude measured from the layer's Dirac point,
and the spinor Hamiltonian acts on an individual layer's sublattice degree-of-freedom.
We choose the coordinate system depicted in Fig.1 for which the decoupled bilayer Hamiltonian is
$|1\rangle  h(\theta/2) \langle 1| + |2\rangle  h(-\theta/2) \langle 2| $ where $|i\rangle\langle i|$
projects onto layer $i$.

We derive a continuum model for the tunneling term by assuming that the inter-layer tunneling amplitude between
$\pi$-orbitals is a smooth function of separation projected onto the graphene planes.
A $\pi$-band tight-binding calculation then yields (see Supplementary Information) a continuum limit
in which tunneling is local.  We find that the  amplitude for an electron residing on sublattice $\beta$ in one layer to hop to sublattice $\alpha$ on the other layer is
\be
T^{\alpha\beta}(\bm{r}) = w \; \sum_{j=1}^{3} \;  \exp(-i \bm{q}_j \cdot \bm{r}) \;\; T_{j}^{\alpha\beta}
\label{continuum}
\ee
where
\be
T_{1} =
\left(
  \begin{array}{cc}
    1 & 1 \\
    1 & 1 \\
  \end{array}
\right),        \nonumber
\ee
\be
T_{2} = e^{-i\bm{\cG}^{(2)'} \cdot \bm{d}}
\left(
  \begin{array}{cc}
    e^{-i\phi} & 1 \\
    e^{i\phi} & e^{-i\phi} \\
  \end{array}
\right),            \nonumber
\ee
\be
T_{3} = e^{-i\bm{\cG}^{(3)'}\cdot\bm{d}}
\left(
  \begin{array}{cc}
    e^{i\phi} & 1 \\
    e^{-i\phi} & e^{i\phi} \\
  \end{array}
\right),                \label{Tlattice}
\ee
$w=t_{k_{\ty D}}/\Omega$ and $\phi=2\pi/3$.  The three $\bm{q}_j$'s in equation (\ref{continuum}), depicted in Fig.1, are
Dirac model momentum transfers that correspond to the three interlayer hopping processes.
For $\bm{d}=0$ and a vanishing twist angle the continuum tunneling matrix is $3w\delta_{\alpha A}\delta_{\beta B}$,
independent of position.  By comparing with the experimentally known\cite{Kuzmenko} electronic structure
of an AB stacked bilayer we set $w \approx 110 \textrm{meV}$.
The spectrum is independent of $\bm{d}$ for $\theta \ne 0$.
(see Supplementary Information). In the following we therefore set $\bm{d}=0$.

The continuum model hopping Hamiltonian captures the local stacking sequence
of the misaligned layers.
At $\bm{r}=0$, for example, only the $\alpha=A$, $\beta=B$ element of $T$ is non-zero because we have chosen our
origin at a Bernal stacking point.  The local lattice registration changes periodically in space with a
pattern of AB points at which only $T_{A,B}$ is not zero, BA points at which only $T_{B,A}$ is non-zero and
AA points at which only $T_{A,A}=T_{B,B}$ is non-zero.  The periodicity is controlled by the
reciprocal lattice vectors $\bm{q}_{b}$ and $\bm{q}_{tr}$. The magnitude of the moir\'e lattice vector is therefore
$\sqrt{3} a /[2\sin(\theta/2)]$, where $a$ is the carbon-carbon distance in graphene\cite{Moire}.
Because translational symmetry in the continuum model is broken only by the spatially periodic
hopping Hamiltonian, we can apply Bloch's theorem to obtain energy bands at any twist angle
$\theta$, independent of whether the underlying lattice is commensurate.
We expect the continuum model to be accurate up to energies of $\sim$1eV and up to angles of $\sim 10^\circ$.
We solve numerically for the moir\' e bands using the plane-wave expansion illustrated in Fig.1.
Convergence is attained by truncating momentum space at lattice vectors of order of $w/\hbar v $.
The dimension of the matrix which must be diagonalized numerically is roughly $ \sim 10 \; \theta^{-2}$ for small
$\theta$ (measured in degrees),
compared to the $\sim 10^{4} \; \theta^{-2}$ matrix dimension of
microscopic tight-binding models\cite{ShallcrossLong,localization}.

Up to a scale factor the moir\' e bands depend on a single parameter $\alpha=w/vk_\theta$.
We have evaluated the moir\' e bands as a function of their Brillouin-zone momentum
$\bm{k}$ for many different twist angles;
results for $\theta=5^\circ, 1.05^\circ $ and $0.5^\circ$ are summarized in Fig.2.
For large twist angles the low energy spectrum is virtually identical to that of an isolated graphene sheet,
except that the velocity is slightly renormalized.  Large inter-layer coupling effects appear only near the
high energy van Hove singularities discussed by Andrei and co-workers\cite{vanHove_Andrei}.
As the twist angle is reduced, the number of bands in a given energy window increases and the
band at the Dirac point narrows.
This narrowing has previously been expected to develop monotonically with decreasing $\theta$.
As illustrated in Fig.2, we instead find that the Dirac-point velocity vanishes already at $\theta \approx 1.05^\circ$,
and that the vanishing velocity is accompanied by a very flat
moir\' e band which contributes a sharp peak to
the Dirac-point density-of-states (DOS).  At smaller twists
the Dirac-point velocity has a non-monotonic dependence on $\theta$,
vanishing repeatedly at the series of magic angles illustrated in Fig.3.

Partial insight into the origin of these behaviors can be achieved by examining the
simplest limit in which the momentum space lattice is truncated at the first honeycomb shell.
Including the sublattice degree of freedom, this gives rise to the Hamiltonian
\be
\cH_k =
\left(
  \begin{array}{cccc}
    h_{\bm{k}}(\theta/2) & T_{b} & T_{tr} & T_{tl} \\
    T_{b}^{\dagger} & h_{\bm{k_b}}(-\theta/2) & 0 & 0 \\
    T_{tr}^{\dagger} & 0 & h_{\bm{k_{tr}}}(-\theta/2) & 0 \\
    T_{tl}^{\dagger} & 0 & 0 & h_{\bm{k_{tl}}}(-\theta/2) \\
  \end{array}
\right),   \label{H8}
\ee
where $\bm{k}$ is in the moir\' e Brillouin-zone, and $\bm{k_j}=\bm{k}+\bm{q}_j$.
This Hamiltonian acts on four two-component spinors; the first ($\Psi_0$)
is at a momentum near the Dirac point of one layer and the other three $\Psi_j$
are at momenta near $\bm{q}_j$ and in the other layer.
The dependence of $h(\theta)$ on angle is parametrically small and can be neglected.
We have numerically verified that this approximation reproduces the velocity with reasonable accuracy down to the first magic angle
(see inset of Fig.3).  It is possible to demonstrate analytically (see Supplementary Information) that this Hamiltonian has
two zero energy states at $\bm{k}=0$, and a Dirac velocity renormalized by
\be
\frac{v^\star}{v} = \frac{1-3\alpha^2}{1+6\alpha^2}.    \label{vStar}
\ee
The denominator in equation (\ref{vStar}) captures the contribution of the $\Psi_j$'s to the normalization of the wave function while
the numerator captures their contribution to the velocity matrix elements.
For small $\alpha$, equation (\ref{vStar}) reduces to the expression $v^\star/v = 1 - 9 \alpha^2$,
first obtained by Santos {\em et al.}\cite{Santos}.
The velocity vanishes at the first magic angle because it is in the process of changing sign.
The eigenstates at the Dirac point are a coherent combination of components in the two layers
that have velocities of opposite sign!

The distribution of the quasiparticle velocity between the two layers implies exotic
transport characteristics for separately contacted layers.
Consider a counterflow geometry
in which the currents in the two layers flow anti-parallel to one another.
The counter-flow velocity
$v_{\ty{CF}}=v(1+3\alpha^2)/(1+6\alpha^2)$
remains finite at the magic angle when the bands flatten and the density of states is enhanced.
A Kubo formula calculation of the counterflow conductivity then yields (see Supplementary Information)
\be
\sigma_{\ty{CF}} = \sigma_0 \lp \frac{v_{\ty{CF}}}{v^\star} \rp^2,  \label{sigma_CF}
\ee
where $\sigma_0 \sim  e^2\epsilon_{\ty F}\tau/\pi$ is the conductivity of an isolated single graphene layer.
As $\theta$ is reduced from a large value towards $1^\circ$, $v^\star$ is reduced and the density-of-states is correspondingly increased.
The counterflow conductivity is enhanced because of an increased density of carriers, which is {\em not}
accompanied by a decrease in counterflow velocity. For a conventional measurement in which the current in the bilayer is
unidirectional $v_{\ty{CF}}$ in expression (\ref{sigma_CF}) is replaced by $v^\star$.
The increase in the DOS is then exactly compensated by the reduction of the renormalized velocity
and the single layer value of the conductivity is regained.

In summary we have formulated a continuum model description of the electronic structure of
rotated graphene bilayers.
The resulting moir\' e band structure can be evaluated at arbitrary twist angles, not only at commensurate values.
We find that the velocity at the Dirac point oscillates with twist angle, vanishing at a series of
magic angles which give rise to large densities-of-states and large counterflow conductivities.
Many properties of the moir\' e bands are still not understood.  For example, although we are able to explain the largest
magic angle analytically, the pattern of magic angles at smaller values of $\theta$ has so far been revealed only numerically.
Additionally the flattening of the entire lowest moir\' e band at $\theta \approx 1.05^\circ$
remains a puzzle.  Interesting new issues arise
when our theory is
extended to graphene stacks containing three or more layers.
Finally, we remark that electron-electron interactions neglected is this work are certain to be
important at magic twist angles in neutral systems and could give rise to counterflow superfluidity\cite{BECbilayers,counterflowSC},
flat-band magnetism\cite{flatBandFerromagnetism}, or other types of ordered states. \\

The authors acknowledge a helfpul conversation with Gene Mele.
This work was supported by Welch Foundation grant F1473 and by the NSF-NRI SWAN program.

\newpage

\begin{figure}[h]
\includegraphics[width=0.8\linewidth]{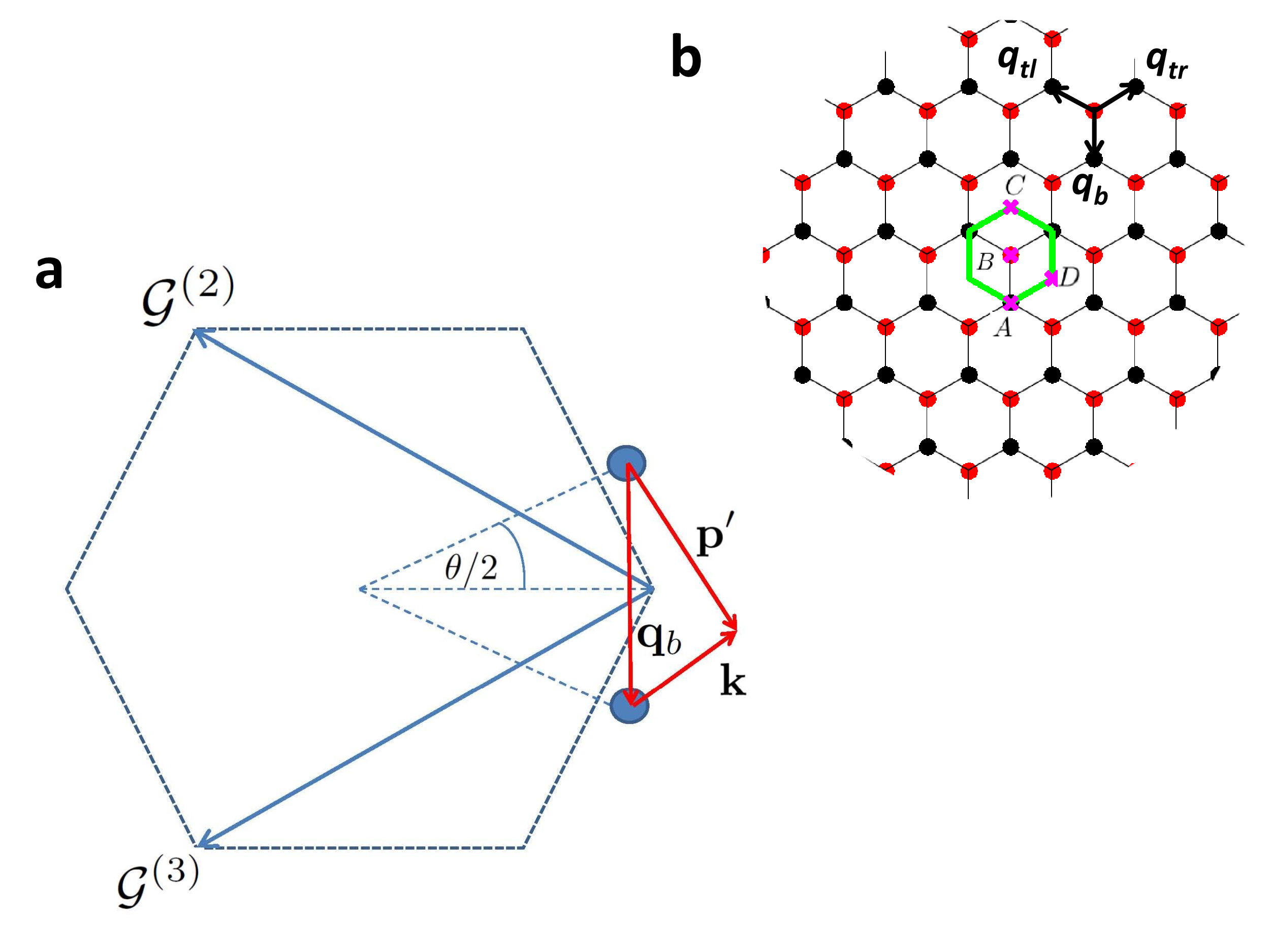}
\caption{
{\bf Momentum space geometry of a twisted bilayer.}  {\bf a)} The circles represent Dirac points of the rotated graphene layers,
separated by $k_{\theta}=2K_{\ty D} \sin(\theta/2)$ where $K_{\ty D}$ is the magnitude of the Brillouin-zone corner wavevector for a
single layer. Conservation of crystal momentum implies that $\bm{p'}=\bm{k+q_b}$ for a tunneling process in the vicinity of the plotted Dirac points.
{\bf b)} The three equivalent Dirac points in the first Brillouin zone result in three distinct hopping processes as explained in the Supplementary Information.
Interference between hopping processes with different wavevectors captures the spatial variation of inter-layer coordination that defines
the moir\' e pattern.
For all the three processes $|\bm{q}_j|=k_{\theta}$; however the hopping directions are
$(0,-1)$ for $j=1$, $(\sqrt{3}/2,1/2)$ for $j=2$ and $(-\sqrt{3}/2,1/2)$ for $j=3$.
Repeated hopping generates a K-space honeycomb lattice. The green solid line marks the moir\' e band Wigner-Seitz cell.
In a repeated zone scheme the red and black circles mark the Dirac points of the two layers.}
\label{fig:kLattice}
\end{figure}

\newpage

\begin{figure*}[ht]
\includegraphics[width=0.8\linewidth]{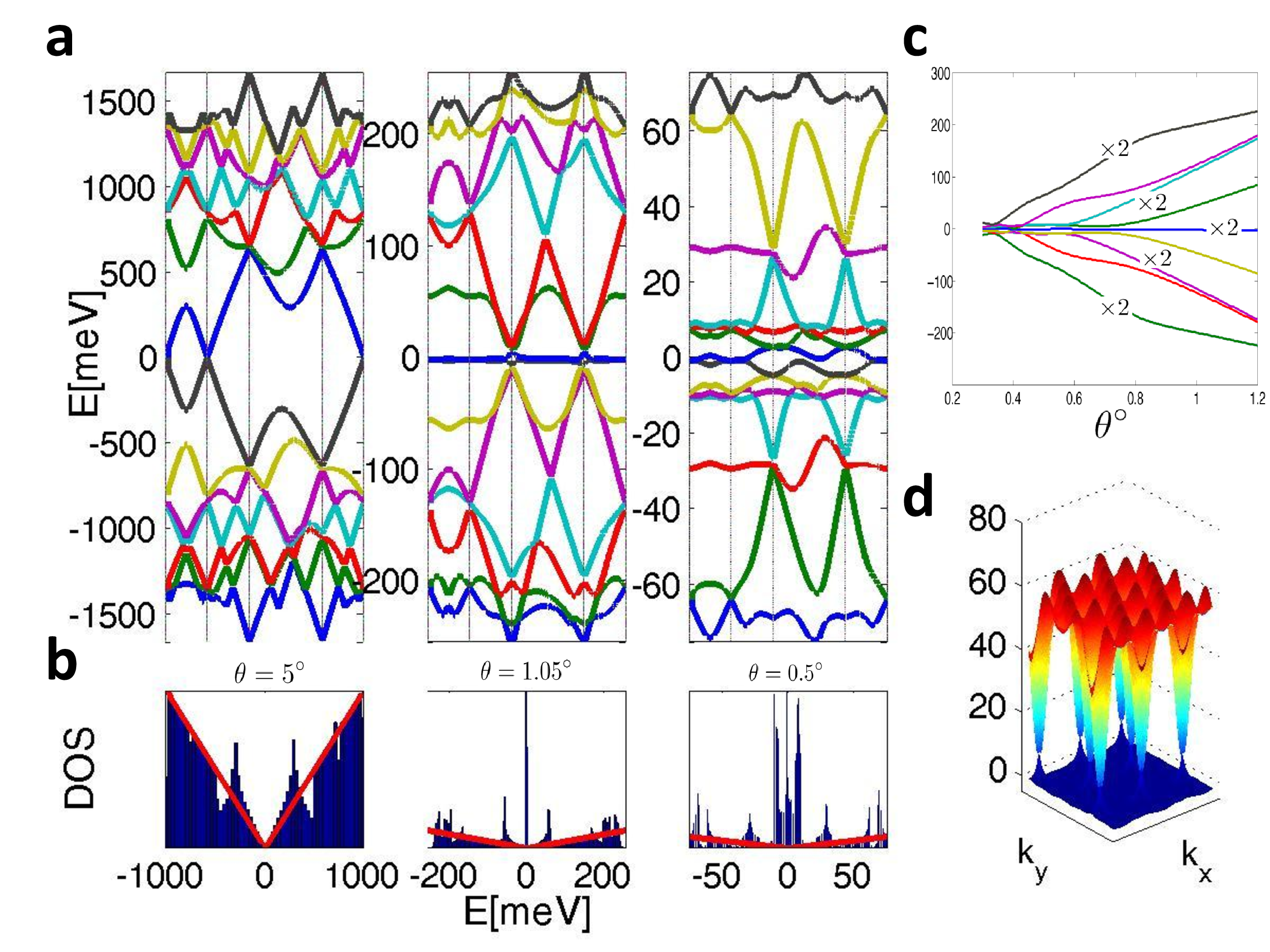}
\caption{{\bf Moire Bands.}  {\bf a)} Energy dispersion for the 14 bands closest to the Dirac point plotted along
the line $A \to B \to C \to D \to A$ (see Fig.\ref{fig:kLattice})
for (left to right) $\theta=5^\circ, 1.05^\circ$ and $0.5^\circ$. {\bf b)} Density-of-states. {\bf c)} Energy as a function of twist angle for the $\bm{k=0}$ states.
Band separation decreases with $\theta$ as also evident from {\bf (a)}.
{\bf d)} Full dispersion of the flat band at $\theta=1.05^\circ$.}
\label{fig:Ebands}
\end{figure*}

\newpage

\begin{figure}[ht]
\includegraphics[width=0.8\linewidth]{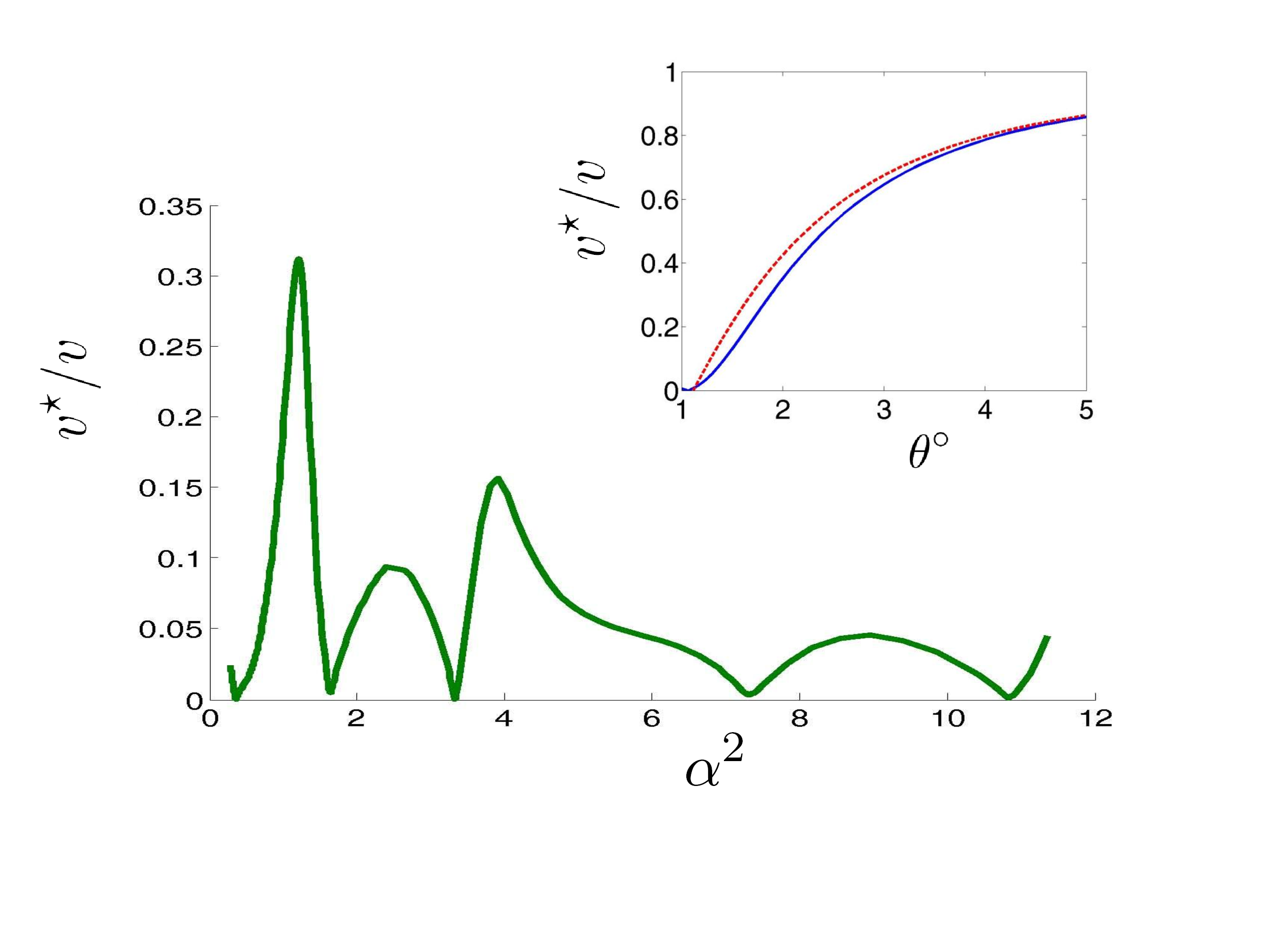}
\caption{ {\bf Renormalized Dirac-point band velocity.} $v^\star$ vs. $\alpha^2$ where $\alpha=w/v k_\theta$ for $0.18^\circ<\theta<1.2^\circ$.
The velocity vanishes for $\theta \approx 1.05^\circ, 0.5^\circ, 0.35^\circ, 0.24^\circ$, and $0.2^\circ$.
Inset: The renormalized velocity at larger twist angles.  The solid line corresponds to numerical results and dashed line corresponds
to analytic results based on the 8-band model.}
\label{fig:Ebands}
\end{figure}

\newpage

\centerline{\Large \bf Supplementary Information}

\section{Tunneling matrix}

The tunneling matrix $T^{\alpha\beta}_{\bm{kp'}}$ describes a process in which
an electron with momentum $\bm{p}'$ residing on sublattice $\beta$ in one layer hops to a momentum state $\bm{k}$ and sublattice $\alpha$ in the
other layer. A tight binding calculation shows that
\be
T^{\alpha\beta}_{\bm{kp'}} = \sum_{\bm{G_1 G_2}} \frac{t_{\bm{\bar{k}+G_1}}}{\Omega}
e^{i [\bm{G_1\tau_{\alpha}} -\bm{G_2 (\tau_{\beta}-\tau)} - \bm{G'_2 d}]}
\delta_{\bm{\bar{k}+G_1,\bar{p}'+G'_2}}
\label{Tkp}
\ee
where $\Omega$ is the unit cell area, $\bm{d}$ linearly displaces one layer relative to the other,
$t_{\bm{q}}$ is the Fourier transform of the tunneling amplitude $t(\bm{R})$, $\tau $ is the vector connecting the two atoms within a unit cell,
$\tau_{\ty A}=0$, $\tau_{\ty B}=\tau$ and
$\bm{\bar{k}}=\bm{K}_{\ty D}+\bm{k}$ with $\bm{K}_{\ty D}$ being the Dirac momentum.
In equation (\ref{Tkp}) we have placed the origin at a point where the $A$ sublattice of one layer
lies above the $B$ sublattice of the other layer when $\bm{d}=0$, the sums are over reciprocal lattice vectors
and the primed vectors $\bm{p'}=M\bm{p}$ and $\bm{G'}_2=M \bm{G}_2$ have been
transformed by the rotation matrix $M$.  Note that the crystal momentum is conserved by
the tunneling process because $t$ depends only on the difference between lattice positions.
A closely related but slightly different expression appears in Ref.\cite{FLGus}
in which we chose the origin at a honeycomb lattice point.  The present convention is more
convenient for the discussion of small rotations relative to the Bernal arrangement.

The continuum model version for $T$ is obtained by measuring wavevectors in both
layers relative to their Dirac points and assuming that the deviations are small compared to
Brillouin-zone dimensions.  $t(q)$ vanishes rapidly with $q$ on
a Brillouin-zone scale (this is  partly because the inter-layer distance is larger than the
honeycomb lattice constant).  Therefore only
values of $\bm{G_1}$ for which $|\bm{K_{\ty D}+G_1}|=|\bm{K_{\ty D}}|$ contribute significantly to $T$.
The three possible values of $\bm{G}_{1}$  in Eq.(\ref{Tkp}) are therefore $\bm{0},\bm{\cG}^{(2)},\bm{\cG}^{(3)}$ where the latter two vectors connect a Dirac point
with its equivalent first Brillioun zone counterparts.  (See Fig.1.) Summing the three terms we find equation (1) in the main text.

\section{Independence of the spectrum on $\bm{d}$}

Here we show that the spectrum of misaligned bilayers is independent of linear translations of one layer with respect to the other
using a unitary transformation that makes the Hamiltonian independent of $\bm{d}$. Consider $H_{\bm{Q}}$ where $\bm{Q}$ is a momentum in the first moir\'e Brillouin zone.
With each momentum on the k-space triangular sublattice (see Fig.1)
\be
\bm{k} = \bm{Q}+n\bm{q_1} + m\bm{q_2}       \nonumber
\ee
where $\bm{q_1}=k_\theta(1/2,\sqrt{3}/2)$, and  $\bm{q_2}=k_\theta(-1/2,\sqrt{3}/2)$ we associate
the phase
\be
\phi_\bm{k} = n\bm{G_2' \cdot d} + m\bm{G_3' \cdot d}.      \nonumber
\ee
The phase associated with momentum $\bm{k-k_\theta\hat{y}}$ on the other sublattice is $\phi_\bm{k}$ as well.
In terms of the new basis states $\exp(i\phi_{\bm{k}})|\bm{k}\alpha\ra$ the Hamiltonian $H_\bm{Q}$
is $\bm{d}$-independent.

\section{Renormalized velocity in the 8-band model}

For twist angles $\theta \gtrsim 2^\circ$ the 8-band Hamiltonian $\cH_{\bm{k}}$, given by Eq.(3) in the main text, adequately accounts for the low energy spectrum.
The renormalized velocity $v^\star = \partial_{\bm{k}} \epsilon^\star_{\bm{k}}|_{\bm{k}=0}$ follows from the spectrum $\epsilon^\star_{\bm{k}}$ of the twisted bilayer.
We find the spectrum perturbatively in $\bm{k}$.
The Hamiltonian is expressed as a sum of the $\bm{k}=0$ term $\cH^{(0)}_{\bm{k}}$ and the k-dependent term
\be
\cH_{\bm{k}}^{(1)} = -v
\left(
  \begin{array}{cccc}
    \sigma \cdot \bm{k} & 0 & 0 & 0 \\
    0 & \sigma \cdot \bm{k} & 0 & 0 \\
    0 & 0 & \sigma \cdot \bm{k} & 0 \\
    0 & 0 & 0 & \sigma \cdot \bm{k} \\
  \end{array}
\right),                 \nonumber
\ee
and solved to leading order in $\cH^{(1)}$.

Consider the $\bm{k=0}$ term in the Hamiltonian.
The wave functions of $\cH^{(0)}$ are expressed as $\Psi=(\Psi_0,\Psi_1,\Psi_2,\Psi_3)$ each $\Psi_j$ being a two component spinor.
We now assume that $\cH^{(0)}$ has zero energy eigenstates and prove our assumption by explicitly finding these states.
The zero energy eigenstates must satisfy
\be
\Psi_j = -h_j^{-1} T_j^\dagger \Psi_0.      \label{psi_j}
\ee
Since
\be
T_j h_j^{-1} T_j^\dagger = 0        \label{hopping_identity}
\ee
the equation for the $\Psi_0$ spinor is
\be
h_0 \Psi_0 = 0          \label{psi0}
\ee
i.e. $\Psi_0$ is one of the two zero energy states $\Psi_0^{(1)},\Psi_0^{(2)}$ of the isolated layer. The two zero energy eigenstates of $\cH^{(0)}$ then follow
from equations (\ref{psi_j}) and (\ref{psi0}).
Given that $|\Psi_0^{(j)}|=1$ the normalization of $\Psi$ is given by
\be
|\Psi|^2 = \Psi_0^{(j)\dagger} \lp I + w^2 \sum_j T_j h_j^{-1\dagger} h_j^{-1}  T_j^\dagger   \rp \Psi_0^{(j)}       \label{psi0_normalization}
\ee
with $I$ being the identity matrix.
For the second term in equation (\ref{psi0_normalization}) we use the fact that $h$ is hermitian and the relations
\bea
h_j^{-1} &=& -h_j/\epsilon_\theta^2  \nonumber \\
h_j^2 &=& \epsilon_\theta^2 I   \label{h2_identity} \\
\sum_j T_j T_j^\dagger &=& 6 I   \nonumber
\eea
to obtain
\be
 w^2 \sum_j T_j h_j^{-1\dagger} h_j^{-1}  T_j^\dagger =  \frac{w^2}{\epsilon_\theta^4} \sum_j T_j h_j^2  T_j^\dagger = 6\alpha^2,    \nonumber
\ee
and $|\Psi|^2=1+6\alpha^2$.

To leading order in $k$ the energies are therefore given by the eigenvalues of
\be
\cH^{(1)}_{ij} = \me{i}{\cH_{\bm{k}}^{(1)}}{j} = -\frac{v}{1+6\alpha^2} \ \Psi_0^{(i)\dagger} \left[ \sigma \cdot \bm{k} + w^2 \sum_j T_j h_j^{-1\dagger} \sigma \cdot \bm{k} h_j^{-1}  T_j^\dagger   \right] \Psi_0^{(j)}.
\nonumber
\ee
Using familiar Pauli matrix identities and equations (\ref{hopping_identity},\ref{h2_identity}) we obtain
\be
w^2 \sum_j T_j h_j^{-1\dagger} \sigma \cdot \bm{k} h_j^{-1}  T_j^\dagger = -\alpha^2 \sum_j T_j \sigma \cdot \bm{k} T_j^\dagger  = -3\alpha^2 \sigma \cdot \bm{k}.    \nonumber
\ee
Aside from a renormalized velocity
\be
\cH^{(1)}_{ij} = -v \frac{1 - 3\alpha^2}{1+6\alpha^2} \Psi_0^{(i)\dagger} \sigma \cdot \bm{k} \Psi_0^{(j)}      \nonumber
\ee
is therefore identical to the continuum model Hamiltonian of single layer graphene, $\epsilon_{\bm{k}}^\star=v^\star k$, where the renormalized velocity is given by equation (4)
in the main text.

\section{Counterflow conductivity in the 8-band model}

We find the counterflow conductivity $\sigma_{\ty{CF}}$ of the bilayer system.
This conductivity relates a counterflow current to the electric fields that induce it; the latter being oppositely orientated in the two layers.
We restrict the calculations to twist angles $\theta \gtrsim 2^\circ$ for which the 8-band model is valid and to the semiclassical regime in which
$\epsilon_{\ty F}\tau > 1$. We assume that only charge carriers in the linearly dispersing sector of the lowest conduction band contribute to the conductivity, i.e. that
the Fermi momentum is much smaller than $k_\theta$ and that $1/\tau_0<\hbar v k_\theta$ where $\tau_0$ is single particle lifetime.

Using the Kubo formula we find that
\be
\sigma_{\ty{CF}} =  \frac{e^2 g}{\pi} \sum_{\bm{k}\mu} |\me{\psi_k}{v^{x}_{\ty{CF}}}{\psi_k}|^2 \left[ Im \{G_{k\mu}^r(\epsilon_F) \} \right]^2
\ee
where $g=4$ accounts for the spin and valley degeneracies,
\be
v^{x}_{\ty{CF}} = -v
\left(
  \begin{array}{cccc}
    \sigma_x & 0 & 0 & 0 \\
    0 & -\sigma_x & 0 & 0 \\
    0 & 0 & -\sigma_x & 0 \\
    0 & 0 & 0 & -\sigma_x \\
  \end{array}
\right)
\ee
is the x-component of the counterflow velocity operator(we set the electric fields along the x-axis),
\be
G^r_{k\mu}(\omega) = \frac{1}{\omega-\epsilon^\star_{k\mu} + i/2\tau_0}
\ee
is the retarded Green function with $\mu$ labeling the two Dirac bands,
and $\epsilon_{k\mu}^\star=\mu v^\star k$ is the energy dispersion of the twisted bilayer at small momenta.
For an electron-doped system the valence band can be neglected and
\be
\sigma_{\ty{CF}} \approx e^2 g \tau \nu^\star(\epsilon_{\ty F})  \int \frac{d\theta_k}{2\pi}  |\me{\psi_{k\mu}}{v^x_{\ty{CF}}}{\psi_{k\mu}}|^2
\ee
where $\nu^\star$ is the density of states of the twisted bilayer. The vertex function
\be
\me{\psi_{k}}{v_{\ty{CF}}^x}{\psi_{k}} = v_{\ty{CF}} \cos\theta_k,
\ee
where $v_{\ty{CF}} = v(1 + 3\alpha^2)/(1 + 6\alpha^2)$ follows directly
from the previous section if we notice the sign differences between the counterflow velocity operator and the normal velocity operator.
The counterflow conductivity given by equation (5) in the main text readily follows.

\end{document}